\documentclass[12pt]{iopart}
\usepackage{iopams}
\usepackage{amsfonts}
\usepackage{amssymb}
\usepackage{setstack}

\begin{document}

\newcommand{\C}{{\mathbb C}}
\newcommand{\R}{{\mathbb R}}
\newcommand{\na}{\nabla}\newcommand{\pa}{\partial}
\newcommand{\CC}{{\cal C}}
\newcommand{\supp}{\mathrm{supp}}
\newcommand{\Ll}{L_{\mbox{\rm\small loc}}}
\newcommand{\Hl}{H_{\mbox{\rm\small loc}}}

\newtheorem{thi}{\hspace*{-1.1mm}}[section]
\newcommand{\bthm}{\begin{thi} {\bf Theorem. }}
\newcommand{\bdef}{\begin{thi} {\bf Definition. }}
\newcommand{\ethi}{\end{thi}}

\newcommand{\todo}[1]{$\clubsuit$\ {\tt #1}\ $\clubsuit$}

\def\sob#1{{\mathcal H}^{#1}}
\def\EPS#1{{#1}_\epsilon{}}
\def\EPSUP#1#2{{#1}^{#2}_\eps{}}
\def\EPSDOWN#1#2{{#1}_{#2}^\eps{}}
\def\geps{\EPS{g}}
\def\eueps{\EPSUP{e}}
\def\gdeps{\EPSDOWN{g}}
\def\gueps{\EPSUP{g}}
\def\ueps{\EPS{u}}
\def\hueps{\hat\EPS{u}}
\def\veps{\EPS{v}}
\def\weps{\EPS{w}}
\def\feps{\EPS{f}}
\def\boxeps{\square^\eps}
\def\hboxeps{\hat\square^\eps}
\def\eps{\epsilon}
\def\mueps{\mu^\eps}

\newcommand{\gs}{\ensuremath{{\mathcal G}} }
\def\sob#1{{\mathcal H}^{#1}}
\def\Re{{\mathbb R}}
\def\css{\subset\!\subset}
\def\Nat{{\mathbb N}}
\def\E{{\mathcal E}}
\def\N{{\mathcal N}}
\def\G{{\mathcal G}}
\def\A{{\mathcal A}}
\def\Sup{{\hbox{sup}}}
\def\D{{\mathcal D}}
\def\d{{\rm d}}
\def\half{{\scriptstyle {1 \over 2}}}
\def\quarter{{\scriptstyle {1 \over 4}}}
\def\p{\partial}
\def\CQG{Class. Quant. Grav. }
\def\cqg{Class. Quant. Grav. }
\def\GRG{J. Gen. Rel. Grav. }
\def\JMP{J. Math. Phys. }
\def\PRD{Phys. Rev. D }

\topical[Generalised Functions in Relativity]
{The use of Generalised Functions and Distributions in General Relativity}
%
\author{R Steinbauer\dag and J A Vickers\ddag}
\address{\dag Department of Mathematics, University of Vienna,
  Nordbergstrasse 15, A-1090 Wien, Austria.\\ Email: 
roland.steinbauer@univie.ac.at\\\quad\\}

\address{\ddag School of Mathematics, University of Southampton,
Southampton, SO17~1BJ, UK.\\ Email: J.A.Vickers@maths.soton.ac.uk}
\begin{abstract}
In this paper we review the extent to which one can use classical
distribution theory in describing solutions of Einstein's equations. We
show that there are a number of physically interesting cases which cannot
be treated using distribution theory but require a more general concept. 
We describe a mathematical theory of nonlinear generalised functions based
on Colombeau algebras and show how this may be applied in general
relativity. We end by discussing the concept of singularity in general
relativity and show that certain solutions with weak singularities may be
regarded as distributional solutions of Einstein's equations.     
\end{abstract}
\submitted{\CQG}
\pacs{0420 -q}
\maketitle

\section{Introduction}\label{intro}

Idealisations play an important role in modelling a wide range of physical
situations. In many field theories a particularly useful idealisation is to
replace an extended source which is concentrated in a small region of space
by a point charge. Such an approximation is physically reasonable provided
the internal structure of the source can be neglected. In a similar way
sources concentrated near a line or a surface can be described in terms of
strings or shells of matter. On trying to describe this idealisation
mathematically it is natural to use a delta function to describe the source,
and hence in this context distributions arise in a natural way.
Distributions are also used to describe a number of other important
physical scenarios such as the description of shock waves and the junction
conditions between matter and vacuum regions.

In the case of linear field theories such as electrodynamics, distribution
theory in fact furnishes a consistent framework which has the following two
important features. Firstly, since the Maxwell equations are linear with
respect to both sources and fields, one can allow distributional solutions
of the field equations as well as classical smooth solutions. Secondly it
is guaranteed that smooth charge densities that are close to those of a
point charge (in the sense of distributional convergence) produce fields
that are close to the Coulomb field.  While the first property permits one
to have a mathematically sound formulation of concentrated sources it is
precisely the latter notion of ``limit consistency'' which renders the 
idealisation physically reasonable.

One would like to have a similar mathematical description of concentrated
sources in the theory of general relativity.  However, general relativity
is different from other field theories in two important respects. Firstly
the Einstein field equations are nonlinear, so that one cannot simply pass
from smooth solutions of the field equations to weak solutions as one can
in a linear theory. More precisely the curvature tensor is a nonlinear 
function of the metric and its first two derivatives. Thus the metric needs 
to be at least $C^2$ to guarantee that the curvature is continuous. 
Mathematically one can weaken this condition to allow a $C^{2-}$ 
(i.e. first derivative Lipschitz continuous) metric and 
this is sufficient for most results in differential geometry to remain
valid. In certain special situations one can lower the differentiability
requirements still further and formulate the field equations in a way which 
avoids ill-defined products of distributions. For example it is possible to 
describe shells of matter \cite{israel1}, \cite{cd} and gravitational
radiation \cite{khan} within the context of classical distribution
theory. Using the standard definition of the curvature
 \begin{equation}
 R^a_{\phantom{a}bcd}=\Gamma^a_{db,c}-\Gamma^a_{cb,d}+\Gamma^a_{cf}\Gamma^f_{db}
 -\Gamma^a_{df}\Gamma^f_{cb}
 \end{equation}
one sees that one wants the connection to have a (weak) derivative and be
locally square integrable in order for the left hand side to make sense as a
distribution. This led Geroch and Traschen \cite{gt} to introduce a class
of ``regular metrics'' for which the components of the curvature are well
defined as a distribution. Geroch and Traschen went on to show that such
regular metrics can only have curvature with singular support on a submanifold of
co-dimension at most one. Thus for a 4-dimensional spacetime a metric
representing a shell of matter could belong to this class but a string or
particle could not.  We will consider such distributional solutions of
Einstein's equations in more detail in \S 2.  The second important respect
in which general relativity is different from other field theories is that
one does not have a fixed background metric, but instead the geometry is
determined by the field equations. Spacetime is described by a manifold
together with a Lorentz metric which is assumed to be sufficiently
differentiable for Einstein's equations to be defined. One then detects the
presence of singularities by showing that the spacetime is incomplete in
some sense. The problem with this approach is that many physically
reasonable spacetimes contain singularities according to this
definition. For example solutions to Einstein's equations representing
cosmic strings are singular. What one would like to do is to
lower the differentiability required of the metric to permit a wide class
of ``distributional geometries'' which represent physically reasonable
solutions and are also mathematically tractable. Owing to the nonlinear
nature of Einstein's equations such distributional geometries will in
general require some nonlinear theory of generalised
functions. Moreover, as pointed out by Isham \cite{isham} distributional 
solutions are not only important in allowing one to deal with concentrated
sources or describing weak singularities but (based on our experience with
linear field theories) should also be included in any path integral
formulation of quantum gravity.

Although there has long been a desire to allow distributional geometries in
general relativity it is only comparatively recently that any real progress
has been made in realising this aim. Many of the early attempts
(e.g. \cite{parker}, \cite{Raju}, \cite{hb1}, \cite{hb2}) 
used methods which were
specifically adapted to the particular problem being considered and whose
general applicability was uncertain.  More recently a number of different
authors (for an overview see \cite{vickersESI}, \cite[Ch. 5.2]{BOOK}) have
investigated distributional geometries using an approach based on Colombeau
algebras.  These were developed by J.F. Colombeau in the 1980's and contain
the smooth functions as a subalgebra and the distributions as a linear
subspace. The key idea of this approach is regularisation through smoothing 
and the use of asymptotic estimates with respect to a
regularisation parameter. In particular, it provides a mathematically
consistent way of multiplying distributions and a unified view on
calculations involving various regularisation procedures.  An important
feature of these algebras is the notion of association which gives a
correspondence between elements of the algebra and distributions. This
allows one to use the power of the algebras to do mathematical calculations
but then use the concept of association to interpret the final result in
terms of classical distributions and give the solutions a physical
interpretation.

In \S2 we review the extent to which it is possible to incorporate
nonlinear operations into classical distribution theory. We will show that
some very limited operations are permitted and that some apparently
reasonable ``multiplication rules'' lead to inconsistencies. We then
consider the implications of this for distributional solutions of
Einstein's equations and end the section by looking in detail at the
properties of the Geroch-Traschen regular metrics. In \S3 we
give a brief review of Colombeau theory and explain how it is able to
circumvent the result of Laurent Schwartz on the impossibility of
multiplying distributions. Historically one of the first singular 
solutions of Einstein's equations to be studied from a distributional 
viewpoint was the Schwarzschild solution which was considered by Parker 
in 1979 \cite{parker}. The key observation is that when written in 
Kerr-Schild coordinates the coefficients of the metric are locally
integrable. The Schwarzschild and Kerr solutions have been
studied in these coordinates by Balasin and Nachbagauer in a number of
papers and in \S4 we  describe their work using the language of the (special)
Colombeau algebra. If one boosts the Schwarzschild solution with velocity $v$
then one obtains, in the limit that the velocity is that of light, the
``ultrarelativistic Schwarzschild solution''. This metric was
investigated by Aichelburg and Sexl \cite{as} who showed that it could be
thought of as a gravitational shock wave. When one does a similar
calculation with the ultrarelativistic limit of the Reissner-Nordstr{\o}m 
solution one obtains a solution with vanishing electromagnetic field but
$\delta$-function energy density. In \S5 we show how Steinbauer was
able to explain this physically surprising result using Colombeau  
algebras. In \S6 we consider geodesic equations for impulsive
gravitational wave spacetimes and show how Colombeau algebras provide
an appropriate way of obtaining the results expected on physical ground
without the need to make use of ad hoc ``rules'' for the multiplication
of distributions. The study of conical singularities is another area
where various authors had used a variety of regularisation procedures
to obtain the physically plausible result that a cone has
$\delta$-function curvature at the vertex. However this result appeared
to be at odds with the results of Geroch and Traschen. 
In \S7 we review the analysis and resolution of the problem by Clarke, 
Vickers and Wilson using the (full) Colombeau algebra. In \S8 we discuss
questions of coordinate invariance of Colombeau algebras and review
work on global and diffeomorphism invariant versions of the construction
leading to a ``nonlinear distributional geometry''. In \S9 we look at 
``generalised hyperbolicity'' and how weak singularities may be regarded 
as distributional solutions of Einstein's equations. Finally we give
some conclusions and an outlook to future work in \S10.

\section{Classical Distribution Theory and General Relativity}

In this section we briefly review the fundamental problems 
encountered when one tries to incorporate nonlinear operations into classical
distribution theory. We will then examine the inherent limitations this
imposes on distributional products and review the consequences for
distributional solutions of Einstein's equations. 

There has been a long history of using generalised function ideas in physics to
model point sources and discontinuous phenomena. Such generalised functions
were put on a sound mathematical footing by the development of the
theory of distributions through the work of S. L. Sobolev \cite{sobolev} 
and L. Schwartz \cite{schwartzBOOK}. 
The basic idea is to make distributions dual to a space of smooth ``test
functions''. To introduce some notation we let $\D(\R^n)$ denote the 
space of smooth functions of compact support on $\R^n$. If $S$ is a linear
form $S: \D(\R^n) \to \C$ then we will denote the action of $S$ on 
$\phi \in \D$ by $\langle S,\phi\rangle$. The vector space of distributions
 $\D'(\R^n)$ is then defined to be the space of continuous linear forms on 
$\D(\R^n)$. In a similar way distributions on an orientable manifold $M$
are defined as continuous linear functionals on the space of compactly 
supported $n$-forms i.e. $\D'(M):=[\Omega^n_c(M)]'$. 
%
A rich theory of
distributional tensor fields (and sections of more general vector bundles) 
has been developed by Marsden in \cite{marsden}; for a pedagogical 
introduction see \cite[Ch.\ 3.1]{BOOK}.

The theory of distributions quickly proved to be extremely successful both 
in applications to the study of linear partial differential equations and in
justifying the use of generalised functions in physics. For example the
Malgrange-Ehrenpreis theorem shows that any linear PDE with  constant coefficients
has a fundamental solution in the space of distributions. However 
the theory soon displayed its natural limitations when in 1957
H. Lewy \cite{lewy} gave his famous example of a linear partial
differential equation with smooth (non-constant) coefficients which 
does not have a distributional solution. A second difficulty with the
theory is that the definition of a distribution as a linear functional does
not make it easy to define the product of distributions. This prevents one
from using distributions to investigate nonlinear PDEs with singular data
or coefficients. 

Although the Lewy  example of a linear PDE without a distributional
solution came as a great surprise, the difficulties that one encounters with
the multiplication of distributions are much easier to understand and can
be seen by looking at some simple cases. In the following we shall briefly 
discuss such examples concentrating on the powers of the Heaviside function 
$H$ and its product with Dirac's delta function $\delta$. If we regard $H$
as a discontinuous function then  $H^m=H$ ($m\in{\mathbb N}$). However if we
differentiate this formula and use the Leibniz rule for the derivative
the following one-line calculation implies the vanishing of the delta function:
\begin{equation}
2H\delta=(H^2)'=(H^3)'=3H^2\delta=3H\delta,\ 
\mbox{so}\ H\delta=0\ \mbox{hence}\ \delta=0.
\end{equation}
Another popular ``multiplication rule'' is
\begin{equation}\label{multrule}
 H\delta={1 \over 2}\ \delta.
\end{equation} 
We demonstrate the problematic nature of this rule (and in fact any rule of
the form $H\delta=c\delta$, $c\in{\mathbb R}$) by considering the simple
ODE 
\begin{equation}
y'(t)=\delta(t)y(t),\qquad y(-\infty)=1,
\end{equation}
(for an amusing discussion of this equation see \cite{hajek}).

Using the ansatz $y(t)=1+\alpha H(t)$ and (\ref{multrule}) we find
$\alpha\delta=(1+\alpha/2)\delta$, hence $\alpha=2$ and the solution takes
the form
\begin{equation}
y(t)=1+2H(t).
\end{equation}
On the other hand a different approach motivated by the requirement of
stability under perturbation is to regularise the singular coefficient by a
sequence $\delta_n$ weakly converging to $\delta$. Then the solution to the
regularised equation is given by
$y_n(t)=\exp(\int_{-\infty}^t\delta_n(s)ds)+1$ which converges to
\begin{equation}
\tilde y(t)=1+(e-1)H(t),
\end{equation}
which obviously does not coincide with the previous solution. 

The deeper reason behind these and all other inconsistencies in this realm
is the incompatibility of the laws of a (commutative) differential algebra
with the formulae $H^m=H$ and $H'\not=0$.  This insight was put into
stringent form by L. Schwartz himself in his incompatibility result
\cite{schwartzIMP}, which says that if the vector space $\D'$ of
distributions is embedded into a differential algebra $(\A,+,\circ)$ then
the following properties are mutually contradictory:
\begin{itemize}
\item[(i)] $\D'$ is linearly embedded into $\A$ and 
$f(x)\equiv1$ is the unity in $\A$.
\item[(ii)] There exist linear derivation operators 
$\pa_i: \A\to\A$ satisfying the Leibniz rule.
\item[(iii)] $\pa_i|_{\D'}$ is the usual partial derivative.
\item[(iv)] $\circ|_{C^k\times C^k}$ (for $k$ finite) is the 
usual pointwise product of functions.
\end{itemize}
It was this result that led to the idea that it was impossible to multiply
distributions. However given that repeatedly differentiating a $C^k$ function
eventually produces a  distribution it is perhaps unreasonable to insist on
(iv) but instead we should only require that the product of smooth functions is the usual
pointwise product. We will see in  \S3 that this is precisely the
condition satisfied by Colombeau algebras.

In the rest of this section we examine the extent to which one can apply
linear distribution theory in the context of general relativity. After
summarising work done in a number of special cases we review a classical
paper by Geroch and Traschen \cite{gt} in which they set up a ``maximal''
distributional framework by finding the ``largest possible'' class of
spacetime metrics which allow for a distributional formulation of the field
equations, and we discuss its limitations.

Spacetimes involving an energy-momentum tensor supported on a hypersurface
of spacetime have long been used in general relativity~(see
\cite{lanczos1,lanczos2,darmois} and~\cite{israel1,obrien}, as well as the
references therein).  Consider a submanifold $S$ of codimension one
dividing spacetime into a ``lower'' and ``upper'' part and let the metric
be smooth up to and including $S$ from each of its sides but allow for a
jump of its first derivatives across $S$. Writing out the Einstein equations in
terms of the extrinsic curvature of $S$ one finds junction conditions
very similar to the ones in electrodynamics (see e.g.~\cite{poisson},
Ch. 3.7). More precisely, the jump of the extrinsic curvature is
interpreted as the surface stress-energy of a surface layer located at
$S$. In the case of $S$ being timelike this arrangement represents a thin
shell of matter, while if $S$ is null it may be interpreted as a thin shell
of radiation (see e.g.~\cite{khan}).  In~\cite{israel1} W. Israel has given
a general formulation of this widely applied approach with the
practical advantage that no reference to any special coordinate system is
required; the four-dimensional coordinates may be chosen freely and hence
may be adapted to possibly different symmetries in the upper and lower part of
spacetime.

On the other hand Lichnerowicz~\cite{lich1}, has given an alternative
description using tensor distributions assuming the existence of an
admissible continuous coordinate system across $S$.  This formalism was
used by Lichnerowicz~\cite{lich2,lich3} and Choquet-Bruhat~\cite{cbWAVES}
to study gravitational shock waves. They derived algebraic conditions on
the metric across the shock (the ``gravitational Rankine-Hugoniot
conditions'') as well as equations governing the propagation of the
discontinuities.  The respective formalisms of Israel and Lichnerowicz were
shown to be equivalent in~\cite{mk}.

The description of gravitational sources supported on two-dimensional
submanifolds of spacetime, however, is more delicate. Israel~\cite{israel2}
has given conditions under which a sensible treatment of the field of a
``thin massive wire'' is possible.  He isolated a class of ``simple line
sources'' which possess a linear energy-momentum tensor and hence allow a
well-defined limit as the wire's radius shrinks to zero. We will look at
line sources in more detail below.

On the other hand Taub~\cite{taub} has claimed to have generalised
Lichnerowicz's formalism to include gravitational sources supported on
submanifolds of arbitrary codimension in spacetime.  However, he had to fix
the ill-defined products by ``multiplication rules'', in particular,
by using equation (\ref{multrule}).

We now begin to review the systematic approach by Geroch and Traschen in
analysing the structure of the nonlinearities of the field equations to see
how far one can get avoiding ill-defined distributional products. 
More precisely, the
quest is for a class of metrics allowing for a distributional formulation
of the Einstein tensor in order to assign to the spacetime---via the field
equations---a distributional energy-momentum tensor representing the
``concentrated'' source. Note that there are two contradictory demands on
this class of metrics: on the one hand these metrics should be ``nice enough''
to permit the distributional calculation of the curvature entities, while
on the other hand they should be ``bad enough'' to have the Einstein tensor
and hence the energy-momentum tensor concentrated on a submanifold of a
high codimension in spacetime.

We write out the coordinate formula of the Riemann curvature tensor in
terms of the Levi-Civita connection and the connection in terms of the metric
\begin{equation}
\label{rt} R_{abc}\,^d\, =\,2
\Gamma^d_{e[b}\Gamma^e_{a]c}+2\pa_{[b}\Gamma^d_{a]c}\,,
\end{equation}
\begin{equation}
\label{rt2} 
\Gamma^a_{bc}
\,=\,g^{ae}(\pa_{(b}g_{c)e}-\frac{1}{2}\pa_c g_{bc})\,\,.  
\end{equation} 
and try to ``save'' these equations by putting just as much restrictions on
the metric tensor as needed to allow for a distributional interpretation of
the respective right hand sides.  For the first term in (\ref{rt}) it is
obviously sufficient to assume $\Gamma^a_{bc}$ to be locally square
integrable. Since $\Ll^2\subseteq\Ll^1$ this requirement actually also
suffices to interpret the second term in (\ref{rt}) as the weak derivative of
the regular distribution $\Gamma^a_{bc}$. Furthermore from equation (\ref{rt2})
we see that it is sufficient to demand $g^{ab}$ to be bounded
locally almost everywhere in order to produce locally square integrable
$\Gamma^a_{bc}$ from locally square integrable first weak derivatives of
$g_{ab}$.  This motivates the following definition.

\bdef A symmetric tensor field $g_{ab}$ on a four-dimensional 
manifold $M$ is called a {\em gt-regular metric} if $g_{ab}$ 
and $g^{ab}\in\Ll^\infty\cap\Hl^1$.
\end{thi} 
In the above definition $\Ll^\infty$ denotes the space of locally bounded
functions and $\Hl^1$ denotes the Sobolev space of functions which are
locally square integrable and also have locally square integrable first
(weak) derivative.  Note that although the above definition appears to be
stronger than that in \cite{gt} it is actually equivalent to the original
one where it was merely demanded that $g_{ab}$ as well as $g^{ab}$
are locally bounded almost everywhere, and the first weak derivatives of
$g_{ab}$ are locally square integrable.  In fact, these assertions imply
$\pa_cg^{ab}\in\Ll^2$ since $g^{ab}=\mbox{cof}(g_{ab})/\det g_{ab}$, where
$\mbox{cof}(g_{ab})$ denotes the cofactor of $g_{ab}$. We also note that
the above conditions will always hold for a $C^{1-}$ metric, for such a
metric will always admit a locally bounded weak derivative.  
For a detailed discussion of the relationship between gt-regularity and
some other regularity conditions in the context of axial 
and cylindrical symmetry we refer to~\cite{wilsonDISS}, chaps. 2.3--2.5.

A further important remark on the notion of gt-regular metrics is in
order. 
While the definition is coordinate invariant with the manifold fixed
beforehand, in the case we are most interested in i.e. when we are dealing
with a singular spacetime in general relativity, the situation is different.
We are not given in advance a coordinate system that includes the
singularity.  So the question of whether a singular metric is gt-regular or
not depends crucially on the choice of the differentiable structure which
is imposed on the manifold to include the singular region.

To see that gt-regular metrics actually allow for the distributional
formulation of Einstein's equations we show that one can build a
distributional  Einstein tensor. 
To do this we need to show that the tensor product of 
the contravariant metric with the Riemann tensor makes sense as a
distribution. By writing the terms
involving  the second derivative as a total derivative we may write this in
the form
\begin{equation}
        g^{ef}R_{abc}\,^d\,
        =\,2g^{ef}\Gamma^d_{m[b}\Gamma^m_{a]c}
        +2\pa_{[b}(g^{ef}\Gamma^d_{a]c})-2
        (\pa_{[b}g^{ef})\Gamma^d_{a]c}\,\,.
\end{equation}
Now the first term involves a product $\Ll^\infty\times\Ll^1$ hence stays
locally integrable. The second term involves a weak derivative of an
$\Ll^1$-tensor field hence may be interpreted as a distribution and
the third term is a product of two locally square integrable fields so is
also locally integrable.

Before discussing convergence for gt-regular
metrics we briefly introduce tensor distributions.
Distributional sections of vector bundles and, in particular, distributional 
tensor fields can be defined as continuous linear forms on suitable test section
spaces but are most easily viewed just as sections with distributional
coefficients,
%
that is
\begin{equation}
{\D'}^r_s(M)=\D'(M)\otimes{\mathcal T}^r_s(M),
\end{equation}
where ${\D'}^r_s$ and ${\mathcal T}^r_s$ denote the spaces of
distributional and smooth $(r,s)$-tensor fields respectively. 

We can now discuss an appropriate notion of convergence for gt-regular
metrics. As already indicated above we would like the Einstein tensor of a
sequence of metrics approximating a gt-regular one to approximate the
Einstein tensor of the gt-regular metric.  The natural notion of
weak convergence for a sequence of locally square integrable tensor fields
$((\mu^{i_1\dots i_r}_{j_1\dots j_s})_n)_n$ is convergence locally in
square integral, i.e.\
\begin{equation}\fl
        (\mu^{i_1\dots i_r}_{j_1\dots j_s})_n\to 0 \quad(n\to\infty)
        \mbox{ iff }
        \int(\mu^{i_1\dots i_r}_{j_1\dots j_s})_n(\mu^{k_1\dots k_r}_{l_1\dots l_s})_n
                \,t^{j_1\dots j_sl_1\dots l_s}_{i_1\dots i_rk_1\dots k_r}
        \to 0\ \in\C
\end{equation}
for all smooth compactly supported $(2s,2r)$-tensor densities $
t^{j_1\dots j_sl_1\dots l_s}_{i_1\dots i_rk_1\dots k_r}$.
Using this notion of convergence one may prove the following theorem.

\bthm {\em(Convergence of gt-regular metrics) }\\
Let $g_{ab}$ and $((g_{ab})_n)_n$ be a gt-regular metric and a sequence of
gt-regular metrics respectively and let
\begin{itemize}
\item[(i)] $((g_{ab})_n)_n$ and $((g^{ab})_n)_n$ be locally uniformly 
bounded, and
\item[(ii)] $(g_{ab})_n\to g_{ab}$, $(g^{ab})_n\to g^{ab}$, and 
$(\pa_a g_{bc})_n\to\pa_a g_{bc}$ locally in square integral.
\end{itemize}
Then $(R_{abc}\,^d)_n\to R_{abc}\,^d$ in ${\D'}^1_3(M)$ and  hence 
$(G_{ab})_n\to G_{ab}$ in ${\D'}^0_2(M)$. 
\ethi
We note that the space of gt-regular metrics is complete
with respect to the notion of convergence defined by hypotheses (i) and (ii).
Moreover, let $g_{ab}$ be a continuous gt-regular metric then there exists
a sequence of smooth metrics $((g_{ab})_n)_n$ converging to $g_{ab}$ in the
sense of (i) and (ii).

Before actually checking which class of gravitational sources may be
described by gt-regular metrics we start with the following
heuristic consideration of the behaviour of gt-regular metrics.  
Suppose $S$ is a $d$ dimensional submanifold of a 4-dimensional spacetime
$M$ and the metric $g_{ab}$ is smooth on $M\setminus S$
but some of its components diverge as one approaches  $S$. 
What order of divergence is allowed if $g_{ab}$ is to be gt-regular? 
Let $r$ be a typical distance from $S$ measured by some background
Riemannian metric $h_{ab}$ and suppose the components of $g_{ab}$ diverge 
at the rate of $r^{-s}$ for some
positive number $s$. Then the weak derivatives of $g_{ab}$ diverge like
$r^{-1-s}$ while the volume element is proportional to $r^{3-d}$. In order
for the derivatives of the metric to be locally square integrable we
therefore require that $2(-s-1)+3-d>-1$, and hence that
\begin{equation}
s<1-\frac{d}{2}\,\,.
\end{equation}
Hence we see that the components of gt-regular metrics must grow more
slowly than a rate of $r^{-1+d/2}$ as one approaches a $d$-dimensional 
submanifold in 4-dimensional spacetime. In particular, the larger the
codimension of the submanifold the more strongly the components of the
metric may diverge. However, as shown by the following theorem, there are
severe constraints on the dimension of $S$. 
\bthm\label{corm} {\em (Concentrated sources from gt-regular metrics) }\\
Let $S$ be a submanifold of dimension $d=(0,1,2,3)$ of a four-dimensional
manifold $M$
and let $T^{i_1\dots i_r}_{j_1\dots j_s}\not=0$ a tensor distribution
satisfying
\begin{itemize}
\item[(i)] $\supp(T^{i_1\dots i_r}_{j_1\dots j_s})\subseteq S$, and
\item[(ii)] $T^{i_1\dots i_r}_{j_1\dots j_s}$ is the sum of a locally 
integrable tensor field and the weak derivative of a locally square
integrable tensor field (hence is of the form of the Riemann tensor of a gt-regular metric).
\end{itemize}
Then $d=3$.
\ethi

This theorem fits the picture described earlier in this section, i.e.\ that
gravitating sources with their support concentrated on a $3$-dimensional
submanifold have been treated successfully in the literature while sources
concentrated on submanifolds of higher codimension have turned out to be
more subtle to deal with. However, it should also be emphasised that although the
gt-regularity conditions are sufficient for the curvature to make sense as
a distribution they are certainly not necessary as shown by the examples
below, so that the above result should not be interpreted as implying that a
spacetime with distributional curvature can only have the source confined
to a 3-dimensional submanifold. Indeed in certain algebraically special 
situations---and using a preferred coordinate
system---some of the curvature quantities may be defined for non
gt-regular metrics. For example impulsive $pp$-waves \cite{penrose1} have
been treated extensively using distributions. 

Summing up, gt-regularity provides us with a large class of ``badly
behaved'' metrics which nevertheless allows one to formulate the field
equations in a ``stable'' way. We are, however, sailing close to the wind
as may be seen from the fact that energy conservation may not be formulated
in general for gt-regular metrics. Indeed, the left hand side of the
Bianchi identities $\na_{[a}R_{bc]de}$ involves a product of the
distributional coefficients of the Riemann tensor with the non-smooth
Christoffel symbols and this is only well defined if one imposes additional
conditions on the metric. Similarly, gt-regular metrics may not be used to
raise or lower the indices of a general tensor distribution since the
tensor product would again involve a multiplication of distributions.

By looking in more detail at the combination of terms that one has in the
expression for the curvature, Garfinkle in \cite{garfinkle} has generalised 
the formalism of Geroch and Traschen to include a slightly more general 
class of metrics. However, in extending the class of metrics in this way
one can no longer establish the convergence theorem which one has for
gt-metrics. One is therefore forced to give up the requirement of 
``limit consistency''. This is another indication that with the Geroch
Traschen definition of regularity one has gone about as far as possible 
using conventional distribution theory. Staying strictly within the 
mathematically and physically consistent setting given by this theory, 
one has to restrict oneself to a class of metrics that excludes physically 
interesting cases such as strings and point particles.
If one wants to describe more general gravitational sources the 
nonlinearity of the field equations forces one  to
go beyond the limits of classical distribution theory and face true 
conceptional problems. A consistent framework
allowing for nonlinear operations on singular (e.g. distributional) objects
is provided by Colombeau's algebras of generalised functions, which
we introduce in the next section.

\section{A brief review of Colombeau theory}
 
In this section we give a brief introduction to Colombeau
algebras. For more details see \cite{colombeau2}, \cite{colombeau3}, 
\cite{biagioni}, \cite{BOOK}. 
As we said in the previous section the definition of distributions as
linear functionals is not well suited to formulate a definition of
multiplication. However it is common to visualise the Dirac delta function as
the limit of a sequence of smooth functions, all with integral one, whose
support gets concentrated at the origin. In fact it is possible to give
these ideas a precise mathematical formulation and an alternative 
sequential approach to distribution theory was developed by Mikusi\' nski
\cite{mik} as early as 1948 (see also Temple \cite{temple}). In this approach
a distribution is an equivalence class of weakly converging 
sequences of smooth functions modulo weak zero-sequences.
Working with a more subtle quotient construction Colombeau was 
able to construct 
a differential algebra $\A$ satisfying conditions (i)--(iii) of \S2 
but with (iv) replaced by
\begin{itemize}
\item[(iv')] $\circ|_{C^\infty\times C^\infty}$ is the 
usual pointwise product of smooth functions.
\end{itemize}
In order to introduce the basic concepts we will start by describing the
special (or simplified) Colombeau algebra on $\R^n$. 
The basic idea is to consider generalised functions as 1-parameter
families of smooth functions $\{f_{\epsilon}\}$. Our basic space will
thus be
\begin{equation}
\E(\Re^n)=\{\{f_{\epsilon}\}: \ 0<\epsilon \leq 1,\ \ f_\epsilon \in C^\infty(\R^n)\}.
\end{equation}
We now want to consider how we can represent a function $f$ of finite
differentiability as an element of this space. If we start with some $\Phi \in
\D(\R^n)$ with integral one then we can rescale this to obtain a 
family of functions
\begin{equation}
\Phi_\epsilon(x)={1 \over {\epsilon^n}}\Phi\left({x \over \epsilon}\right)
\end{equation}
with the property that $\Phi_\epsilon \to \delta$ in $\D'$ as $\epsilon \to
0$. Hence if we take the convolution of $f$ with $\Phi_\epsilon$ we obtain a family
\begin{equation}
\label{smoothmetric}
 f_\epsilon(x)={1 \over {\epsilon^n}}\int f(y) 
\Phi\left({{y-x}\over {\epsilon}}\right) \d^ny.
\end{equation}
of smooth functions that converge to $f$ in $\D'$ as $\epsilon$ tends
to zero. We will refer to  $\Phi$  in the above expression as 
a molifier.

Of course we can also apply the above formula to a smooth function
$f$. However for a smooth function we can also represent $f$ as an element
of $\E$ by considering the constant family $f_\epsilon(x)=f(x)$.
For the case of a smooth function we would like both these possible 
representations to be equivalent. Using a Taylor series expansion to
compare the difference between these expressions we are lead
to define two representations to be equivalent if they differ by a
``negligible function'' which is defined as a 1-parameter family 
of functions which on any compact set vanishes faster than any given 
positive power of $\epsilon$. 
Since we are trying to construct a {\it differential algebra} 
we also require the derivatives of $f$ to  satisfy this
property and the resulting set $\N$ to be an ideal. Clearly $\N$
is not an ideal in $\E(\Re^n)$, but by restricting this space to
``moderate functions'' $\E_M(\Re^n)$ which grow no faster than some
inverse power of $\epsilon$, one does have an ideal and we may define
the differential algebra $\G$ as the quotient. 
 
\bdef
\begin{itemize}
\item [(i)] (Moderate functions)
 \begin{eqnarray*}
 \E_M(\Re^n)&:=&\big\{\{f_\epsilon\}:\  \forall K \css\Re^n, \forall \alpha \in
 \Nat^n_0, \exists p\in\Nat 
 \hbox{ such that} \\
 &&\hphantom{\{\{f_\epsilon\}:\ }
 \sup\limits_{x\in K}|D^\alpha f_\epsilon(x)| \leqslant O(\epsilon^{-p}) \hbox{ as }
 \epsilon\to 0\big\}.\\
 \end{eqnarray*}
\item[(ii)] (Negligible functions)
 \begin{eqnarray*}
 \N(\Re^n)&:=&\big\{\{f_\epsilon\}:\  \forall K \css\Re^n, \forall \alpha \in
 \Nat^n_0, \forall q\in\Nat 
 \hbox{ such that}\\
 &&\hphantom{\{\{f_\epsilon\}:\ }
 \sup\limits_{x\in K}|D^\alpha f_\epsilon(x)| \leqslant O(\epsilon^{q}) \hbox{ as }
 \epsilon\to 0\big\}.\\
 \end{eqnarray*}
\item[(iii)] (Special algebra)
 $$
 \G(\Re^n):=\E_M(\Re^n)/\N(\Re^n)
 $$
\end{itemize}
\end{thi}
Note $K \css \Re^n$ indicates that $K$ is {\it compact} and we have
also employed the standard multi-index notation for $D^\alpha f$.
 
Thus a nonlinear generalised function $f$ denoted by $f=[\{f_\eps\}]$ is an 
equivalence class of moderate sequences of smooth functions modulo negligible 
ones; it is represented by a moderate sequence of smooth 
functions $\{f_\eps\}$. The space $\E_M(\Re^n)$ is a differential 
algebra with pointwise operations and
 since the space of negligible functions is a differential ideal, $\G$
 is also a commutative differential algebra. 

The vector space of distributions is now embedded into the algebra
$\G$ by convolution with a molifier: More precisely we choose
a molifier $\Phi$ which for technical reasons (not to be discussed here)
is a Schwartz function and has all moments  vanishing i.e. 
$\int\Phi(x)x^\alpha\,dx=0\ \forall |\alpha|\geq 1$. Then we embed $T\in\D'$ with
compact support as
\begin{equation}\label{emb}\iota(T)=[\{T*\Phi_\eps\}].\end{equation}
Distributions which are not compactly supported are embedded via a
localised version of (\ref{emb}) using a standard sheaf theoretic construction. 

As remarked earlier one of
 the advantages of the Colombeau approach is that one may frequently
 interpret the results in terms of distributions using the concept of
 association or weak equivalence. A generalised function $f$  
 is said to be associated to a distribution $T \in \D'$ if
 for one (hence any) representative  $\{f_\epsilon\}$ we have
 \begin{equation}
 \forall \phi \in \D, \quad \lim_{\epsilon \to 0}\int
 f_\epsilon(x)\phi(x)\d^nx=\langle T,\phi\rangle
 \end{equation}
 and we then write $f \approx T$.
 Note that not all elements of $\G$ are associated to distributions. 
 
 More generally we say for two generalised functions
 $f \approx g$ if
 \begin{equation}
 \forall \phi \in \D, \quad \lim_{\epsilon \to 0}\int
 (f_\epsilon(x)-g_\epsilon(x))\phi(x)\d^nx=0 
 \end{equation}
 for one (hence any) pair of representatives $\{f_\epsilon\}$, $\{g_\epsilon\}$. 
 Association is an equivalence relation which respects addition and
 differentiation. It also respects multiplication by {\it smooth} functions
 but by the Schwartz impossibility results cannot respect multiplication
 in general.
 
The algebra presented above is the simplest of the Colombeau
algebras and can be readily generalised to arbitrary manifolds
(see \S8). However it does suffer from the disadvantage that the 
embedding $\iota$ of distributions and of functions of finite 
differentiability is not canonical but depends on the choice of molifier 
$\Phi$ (see above).  Thus one has to appeal to mathematical or physical 
arguments outside the theory to justify a particular representation 
of a non-smooth function. We discuss these matters in \S8 where
we also present the construction of so-called full Colombeau
algebras which do posses a canonical embedding of distributions.
In the following sections however we will discuss applications of
Colombeau algebras to general relativity using the language of
the special version. 

\section{The Schwarzschild and Kerr Spacetimes}
 
In this section we review (linear and nonlinear) distributional
treatments of the Schwarzschild and Kerr spacetimes. We use the language 
of the special Colombeau algebra although strictly speaking we should be 
using the special version of the theory of generalised tensor 
fields on manifolds, which we will introduce in \S8. However the precise
details will not be needed as we aim at presenting the main 
ideas and concepts in the most elementary way.

Balasin and Nachbagauer considered rotating, charged,  Kerr-Newman 
black-hole solutions in a number of papers (\cite{hb1,hb2,hb3,hb4,hb5,hb}). 
The solutions considered have the feature that they are all 
examples of Kerr-Schild geometries. Such metrics may be written in the form
\begin{equation}
 g_{ab}=\eta_{ab}+fk_ak_b
\end{equation}
 where $\eta_{ab}$ is the flat Minkowski metric, $f$ is an arbitrary
 function, $k_a=\eta_{ab}k^b$ and $k^a$ is null and geodetic with
 respect to $\eta_{ab}$.
 
 The simplest example of such a solution is the Schwarzschild solution
 which in the standard Minkowski coordinates has Kerr-Schild form given
 by
 \begin{equation}
 f={2m \over r}, \quad k^a=(1,x^i).
 \end{equation}
 The expression for the Ricci tensor of a Kerr-Schild metric takes the
 surprisingly simple form
 \begin{equation}
 R^a_b=\half\eta^{cd}\eta^{ea}[\p_e\p_c(fk_dk_b)+\p_b\p_c(fk_dk_e)
 +\p_c\p_d(fk_ek_b)].
 \end{equation}
 The energy momentum tensor is then given by Einstein's equations as
 \begin{equation}
 T^a_b=R^a_b-\half\delta^a_bR.
 \end{equation}
 One may then calculate the energy momentum tensor as follows. The
 Kerr-Schild form is regarded as being valid on the {\it whole} of
 Minkowski space with ${2m \over r}$ replaced by some suitable
 regularised function.
 One now considers this as an element of $\G$ and computes the
 components of $T^a_b$ (in Minkowski coordinates) as elements of
 $\G$. Finally one can show that
 \begin{eqnarray}
 T^0_0 &\!\approx&\! -m\delta^{(3)} \label{1}\\
 T^a_b &\!\approx&\! 0 \quad \hbox{otherwise.} \label{2}
 \end{eqnarray}
 
Different authors (\cite{parker,sakane,para}) using various regularization
procedures have also assigned a distributional energy-momentum tensor to
the Schwarzschild geometry. These approaches have been compared using
the language of the special algebra in \cite{sscol}.

 The calculation of the energy-momentum tensor for the Kerr solution is
 significantly harder. Unlike the Schwarzschild case we cannot easily
 write the Kerr solution as the limit of a one parameter family of
 regular Kerr-Schild metrics. The problem arises because of the
 topology of the maximal analytic extension of the Kerr solution which
 leads to a branch singularity when the metric is written using the
 standard ``flat'' Kerr-Schild decomposition. Balasin  \cite{hb5}
 avoided this problem by considering metrics of generalised Kerr-Schild
 form. These are metrics which can be written
 \begin{equation} 
 g_{ab}={\hat g}_{ab}+fk_ak_b
 \end{equation}
 where ${\hat g}_{ab}$ is now a background metric, $k_a={\hat
 g}_{ab}k^b$ and $k^a$ is null and geodetic with respect to ${\hat g}_{ab}$
 (and also $g_{ab}$ because of the form of the metric).
 One now has
 \begin{equation}
 \fl
R^a_b={\hat R}^a_b+{\hat g}^{cd}{\hat g}^{ef}
{\hat R}^a_{\phantom{a}ceb}fk_dk_f+\half{\hat g}^{cd}{\hat g}^{ea}
 [\p_e\p_c(fk_dk_b)+\p_b\p_c(fk_dk_e)+\p_c\p_d(fk_ek_b)]
 \end{equation}
 where ${\hat R}^a_{\phantom{a}bcd}$ is the curvature of ${\hat g}_{ab}$.
 
 We may write the Kerr metric in generalised Kerr-Schild form by taking
 the background metric to have the form
 \begin{equation}
{\hat g}_{ab}dx^adx^b=-dt^2+{\Sigma \over {r^2+a^2}}dr^2+\Sigma d\theta^2+
 (r^2+a^2)d\phi^2
 \end{equation}
 where $\Sigma=r^2+a^2\cos^2\theta$ and $(t,r,\theta,\phi) \in
 \Re^2\times S^2$. 
 Performing all the calculations in the special algebra one may
 compute $\sqrt{\hat g}R^a_b$ which is found to have the following
 associated distribution
 \begin{equation}
\fl
 \sqrt{\hat g}R^a_b \approx
 2\pi\delta(\cos\theta)(-a\delta(u)\p^a_u du_b+\p^a_\theta d\theta_b 
 +m\delta'(u)(\p^a_t-(1/a)\p^a_\phi)(dt_b+ad\phi_b).
 \end{equation}
 
 The above calculation was carried out by Balasin using the special
 algebra. This is probably the only practical way of doing the
 calculation given the topology of the manifold and the complexity of
 the metric. It suffers from the usual problem when using the
 special algebra of a non-canonical embedding. Rather than embed
 using a convolution, the embedding has been chosen to preserve the
 generalised Kerr-Schild form. However the embedding used is not unique 
 within this class and it would be desirable to show that any
 reasonable embedding which preserved the form of the decomposition
 gave the same result. It is even less clear that a ``natural
 regularisation'' in some other coordinate system would give the same
 result. Nevertheless the calculation is an impressive example of the
 complicated calculations that can be performed in general relativity using 
 the special algebra.

\section{Ultrarelativistic Black Holes}
 
 In 1971 Aichelburg and Sexl \cite{as} derived the ultrarelativistic 
 limit of the Schwarzschild geometry. Below we give a description of 
 the limit using the language of the special algebra. 
 We start by considering the Kerr-Schild form of the Schwarzschild
 metric written in double null coordinates $u=t-x$ and $w=t=x$,
 \begin{equation}
 ds^2=dudw-dy^2-dz^2+{2m \over r}k_ak_bdx^adx^b \label{3}
 \end{equation}
 where in these coordinates $k^a=((r-x)/r,(r+x)/r,y/r,z/r)$.
 
 The Minkowski background enables us to have a well defined concept of
 boost and we may therefore boost the solution by velocity $v$ along
 the $x$-axis. We therefore write
 \begin{equation}
 u=\sqrt{{{1+v} \over {1-v}}}{\bar u} \qquad 
 w=\sqrt{{{1-v} \over {1+v}}}{\bar w}
 \end{equation}
 and to keep the energy of the ``particle'' finite we rescale the mass
 according to the special relativistic formula
 \begin{equation}
 m=(1-v^2)^{\half}p.
 \end{equation}
 Substituting this into (\ref{3}) gives a 1-parameter family of metrics
 depending on the boost velocity $v$. We are interested in the
 ultrarelativistic limit in which $v$ reaches the speed of light (i.e.\ $v
 \to 1$), so we replace $v$ by $1-\epsilon$ and regard 
$\tilde g_{ab}:=[g_{(\epsilon)ab}]$
 as an element of the special algebra. It is readily shown that most
 of the terms in the perturbation ${2m \over r}k_ak_b$ are associated
 to zero. The only surviving term is ${2m \over r}k_0k_0$. Using
 calculations very similar to Steinbauer \cite{rn} we find this is
 associated to $8p\ln\rho\delta(u)$ and hence $\tilde g_{ab} \approx
 g_{(0)ab}$ where
 \begin{equation}
 g_{(0)ab}dx^adx^b=8p\ln\rho\delta(u)du^2+dudw-dy^2-dz^2.
 \end{equation}
 This metric describes a pp-wave and is flat everywhere except on the
 null plane $u=0$ which contains the ``particle''.
 This line element was first derived by
 Aichelburg and Sexl \cite{as}, who started with Schwarzschild written in
 isotropic coordinates and simultaneously boosted the solution and made
 a $v$-dependent coordinate transformation to compute the limiting
 metric.
 
 It should be pointed out that this result is entirely consistent with
 the calculation of the energy-momentum tensor of the Schwarzschild
 solution given in the previous section. The ultrarelativistic limit of
 the latter in the $(u,w,x,y)$ coordinate system is associated to
 \begin{equation} 
 \delta(u)\delta^{(2)}(y,z)p_ap^b \quad \hbox{where } p_a=(1,0,0,0),
 \end{equation}
 which is just the energy-momentum tensor of $\tilde g_{ab}$. 
 Indeed this observation was used by Balasin and Nachbagauer \cite{hb3} to {\it derive}
 the ultrarelativistic limit of the Schwarzschild and Kerr geometries (see also \cite{baho1}).
 
 It is also possible to calculate the ultrarelativistic limit of the
 Reissner-Nordstr{\o}m solution. However in this case it is also
 necessary to rescale the charge according to the formula
 \begin{equation}
 e^2=(1-v^2)^\half f^2, \quad (f \hbox{ a constant})
 \end{equation}
in order to obtain a distributional limit at all as $v$ tends to the speed of light. 
The limiting metric was found by Loust\'o and S\'anchez \cite{ls-rn} using
the methods of \cite{as} to be
 \begin{equation}
 ds^2=\left\{8p\ln\rho+{{3\pi f^2} \over {2\rho}}\right\}\delta(u)du^2
 +dudw-dy^2-dz^2. \label{**}
 \end{equation}
 This was confirmed using a calculation in $\G$ by Steinbauer \cite{steinDIPL}.
 The solution obtained again represents a pp-wave and is flat everywhere
 except on the null plane $u=0$. The ultrarelativistic limit of the
 electromagnetic energy-momentum tensor of Reissner-Nordstr{\o}m 
 is also found to be
 compatible with the energy-momentum tensor of (\ref{**}).

However the rescaling of the charge has the rather unexpected effect that
while the electromagnetic field vanishes in the $\D'$-limit the
ultrarelativistic energy-momentum tensor does not (cf
\cite{ls-rn}). Steinbauer in \cite{rn} rephrased this fact using
association in $\G$ i.e. showed that all the components of the
electromagnetic field were associated to zero while the $00$-component of
the energy-momentum tensor was associated with a multiple of the delta
function. However it must be stressed that regarded as elements of $\G$ the
electromagnetic field components are {\it non-zero} (even though they are
{\it associated} to zero). Calculating the energy-momentum tensor of this
field within $\G$ gives the correct result. There is nothing unusual within
Colombeau theory in having objects in $\G$ which are associated to zero
having products which are not associated to zero. This is simply a
reflection of the fact that association does not respect multiplication in
general.

The procedure of Aichelburg and Sexl has also been used to
derive the ultrarelativistic limit of the Kerr metric
by several authors see \cite{ls-kn,ls-spin,fp,hs}.
In fact a number of  different sources such as  cosmic strings, 
domain walls and monopoles
have been boosted to obtain ultrarelativistic spacetimes of impulsive pp-waves
which in turn have been used to describe (quantum) 
scattering processes of highly energetic particles (see~\cite{v1,v2} 
for an overview).

On the other hand Dray and {}'t\,Hooft \cite{dt} have generalised Penrose's 
``scissors and paste'' method \cite{penrose1} (see also \S6)
to non-flat backgrounds and used it as an alternative way to derive
the Aichelburg-Sexl geometry as well as more general gravitational shock 
waves. Using this method Dray and {}'t\,Hooft derived the spherical shock wave 
due to a massless particle moving at the speed of
light along the horizon of a Schwarzschild black hole which was used
to study the influence of matter, falling into the black hole on its
Hawking-radiation. These ideas lie at the heart of {}'t\,Hooft's S-matrix
approach to quantum gravity~\cite{th}.

\section{Geodesics for impulsive gravitational wave spacetimes}
 
In the previous section we showed how the ultrarelativistic limit of the
Schwarzschild solution lead to an impulsive gravitational wave
spacetime. We now consider geodesics in such spacetimes with the 
metric taking the form
\begin{equation}\label{pp}
ds^2=f(x^A)\delta(u)du^2+dudw-\delta_{AB}dx^Adx^B
\end{equation}
where $A,\ B=2,\ 3$ denote the transverse coordinates. 
These spacetimes have been constructed by Penrose
using his vivid "scissors and paste" approach (see
\cite{penrose2}).
Geodesics for these impulsive gravitational wave spacetimes have been
considered by  Ferrari, Pendenza and Veneziano \cite{fvp}, 
Balasin \cite{hb-geo} and Steinbauer \cite{geo}. As one might expect, 
one can regularise the geodesic equations, and show that the geodesics 
consist of broken and refracted straight lines.
 
A more rigorous derivation of these results using existence and uniqueness
theorems within the Colombeau algebra has been given by Kunzinger and
Steinbauer in \cite{geo2}.  More precisely they replaced the delta function
in (\ref{pp}) by a generalised function $D$ possessing a so called strict
delta net $\{\rho_\eps\}$ as a representative i.e. they considered the
generalised line element
\begin{equation}\label{ppg}
 \hat ds^2=f(x^A)D(u)du^2+dudw-\delta_{AB}dx^Adx^B
 \end{equation}
with $D=[\{\rho_\eps\}]$ and $\mbox{\rm supp}(\rho_\eps) \to \{0\}$ and
$\int \rho_\eps (x) \,dx \to 1$ as $\eps\to 0$ and $||\rho_\eps||_{L^1}$
uniformly bounded in $\eps$.  They were able to show that the geodesic as
well as the geodesic deviation equation for the metric (\ref{ppg}) may be
solved uniquely in $\G$.  Moreover these unique generalised solutions
possess the physically expected associated distributions. Note that
diffeomorphism invariance of these results is assured by diffeomorphism
invariance of the class of strict delta nets. Note further that strictly
speaking these calculations have been performed using the concept of
generalised functions taking values in a manifold, cf \S8, since
geodesics are curves from an interval into spacetime.

In the literature impulsive pp-waves have frequently been described in
different coordinates where the metric tensor is actually continuous (cf
\cite{penrose2}). In the special case of a plane wave,
$f(x,y)=x^2-y^2$ and $u_+$ denoting the kink function,
\begin{equation}\label{cm}
        ds^2\,=\,(1+u_+)^2dX^2+(1-u_+)^2dY^2-dudV\,.
\end{equation}    
Clearly a transformation relating (\ref{cm}) and (\ref{pp}) cannot even be
continuous, hence in addition to involving ill-defined products of
distributions it changes the topological structure of the
manifold. However, the two mathematically distinct spacetimes are
equivalent from a physical point of view, i.e.\ the geodesics and the
particle motion agree on a heuristic level (see \cite{bis-proc}).

Using their results on the geodesic equation in $\G$, Kunzinger and
Steinbauer in \cite{trsf} succeeded in showing that the discontinuous
change of coordinates is just the associated distributional map of a
generalised coordinate transform.  More precisely modelling the
(distributional form of the) impulsive pp-wave metric in a diffeomorphism
invariant way by the generalised metric (\ref{ppg}) the latter may be
subject to a generalised change of coordinates $T$. In either coordinates
the associated distributional metric may be computed giving the
distributional (respectively  the continuous) 
form of the pp-wave metric. Physically
speaking the two forms of the impulsive metric arise as the
(distributional) limits of a sandwich wave in different coordinate systems.
Hence impulsive pp-waves are indeed sensibly modelled by the generalised
spacetime metric (\ref{ppg}). However in the 
different coordinate systems a different distributional picture is obtained.

A similar situation arises in the case of impulsive spherical waves which
have also been introduced in \cite{penrose2}. In this case the
distributional form of the metric arises as an impulsive limit
of type-N Robinson-Trautman solutions, which however due to the
fact that the metric is quadratic in the profile function formally
involves the square of the delta function. An explicit discontinuous 
coordinate transformation relating this form of the metric with the continuous
form was given in \cite{jiri-exp}. A study of this situation using
Colombeau methods relies on a better understanding of the geodesics
of these spacetimes; a study of these has been initiated in
\cite{geo-exp}.

\section{Conical Singularities and Cosmic Strings}
 
 An important example of a two dimensional singularity 
 is provided by the conical spacetime
 \begin{equation}
 ds^2=dt^2-dr^2-A^2r^2d\phi^2-dz^2.
 \end{equation}
 where $A \neq 1$ and $\phi$ is a standard $2\pi$-periodic angular coordinate.
 This spacetime is locally flat for $r \neq 0$ and heuristic arguments
 suggest that it has delta-function like curvature on the $r=0$ axis.
 The corresponding energy-momentum tensor then describes a string with
 stress equal to the mass per unit length $\mu$, where
 $\mu=2\pi(1-A)$. This is precisely the form of the energy-momentum
 tensor of a cosmic string in the weak field thin string limit. Such
 spacetimes are also important mathematically as they provide  simple
 examples of quasi-regular singularities (i.e.\ singularities for which
 the components of the Riemann tensor measured in a parallely propagated
 frame tend to a well defined limit, see e.g.\ Vickers \cite{Vickers} for further
 details).
 
 Unfortunately such spacetimes are not gt-regular, 
 so one cannot expect the curvature to
 be well-defined using conventional distribution theory. Furthermore
 Geroch and Traschen showed how it was possible to obtain different
 values of the mass per unit length by taking different
 regularisations. Because of this Clarke, 
 Vickers and Wilson \cite{cvw} chose to
 investigate the curvature of the cone using the ``full'' Colombeau
 algebra where one has a {\it canonical} embedding of
 distributions. See \S8 below for further details.
 
 The singular part of the curvature arises from the conical singularity
 in the 2-cone
 \begin{equation}
 ds^2=dr^2+A^2r^2d\phi^2
 \end{equation}
 and for simplicity we will describe the calculation of the curvature
 (density) of this metric.  Because polar coordinates are not well defined 
 at the origin one must first write 
 the metric in a regular coordinate system which includes the origin. 
 To make things simple we will choose to work in Cartesian coordinates 
 but as will be shown below the final result is independent of the coordinates used.
 In Cartesian coordinates one has
 \begin{equation}
 g_{ab}=\half(1+A^2)\delta_{ab}+\half(1-A^2)h_{ab},
 \end{equation}
 where 
 \begin{equation}
 h_{ab}=\left(\begin{array}{cc}
 {{x^2-y^2} \over {x^2+y^2}} & {{2xy} \over {x^2+y^2}} \\
 {{2xy} \over {x^2+y^2}} & {{y^2-x^2} \over {x^2+y^2}} 
 \end{array}\right).
 \end{equation}
 We now regard the metric as an element of $\E_M(\Re^2)$ by taking the
 convolution of the components $g_{ab}$ with an arbitrary molifier.
 To  find these we need to calculate
 \begin{equation}
 \tilde h_\epsilon(x,y)={1 \over {\epsilon^2}}\int_{\Re^2}
 h(u+x,v+y)\Phi((u/\epsilon),(v/\epsilon))\d u\d v
 \end{equation}
 where $ h(x,y):=e^{2i\phi}$.
 
 By expanding in circular harmonics, using the compactness
 of the support of the molifier and using the residue theorem one can
 obtain a Fourier series for $\tilde h$ and hence for 
 ${\tilde g}_{(\epsilon)ab}$. 
 One may now estimate the curvature $\tilde R_\epsilon$ of 
 ${\tilde g}_{(\epsilon)ab}$, and use the Gauss-Bonnet theorem to show
 that
 \begin{equation}
 [{\tilde R} \sqrt{\tilde g} ] \approx 4\pi(1-A)\delta^{(2)}.
 \end{equation}
 
 It is important to stress that the methods described here may be used
 to calculate the curvature of the full four dimensional cone, although
 one may no longer use the Gauss-Bonnet theorem to calculate the
 curvature and therefore requires more delicate estimates (see Wilson
 \cite{wilsonDISS} for details). One then obtains the heuristically expected
 energy-momentum tensor. Furthermore it is not hard to modify the
 results to deal with a metric which is not exactly conical, but
 approaches one quadratically as $r \to 0$. A second point to note is
 that the result does not depend upon the coordinates in which the
 calculation is carried out so long as they are smoothly related to Cartesians.
 This may be shown by explicitly transforming to new coordinates
 $X=X(x,y)$ and $Y=Y(x,y)$ and doing the whole calculation in the new
 coordinates. Again more delicate estimates are required but one can
 show that the resulting curvature density transforms in exactly the
 same way as a delta-function (see Vickers and Wilson \cite{invc} 
 for details).
 
 The above calculation shows that the curvature of a 4-dimensional cone,
 when calculated in the full Colombeau algebra, has a curvature which is
 associated to a delta-function which gives a mass per unit length equal to
 the deficit angle as one would expect from the heuristic calculation.
 Note that this result does not say that the curvature is equal to a delta
 function, but simply that it is associated to a delta function.  Thus if
 one works at the level of association this result shows that
 regularisations based upon smoothing convolutions give an unambiguous
 answer for the curvature of the cone. Indeed a number of authors
 have applied different regularisations and obtained the same
 result. Balasin and Nachbagauer \cite{hb1} used a regularisation based on
 a family of smooth hyperboloidal surfaces converging to a cone. A number
 of authors including Marder \cite{marder} and Geroch and Traschen
 \cite{gt} have looked at ``rounding off'' the point of a cone with a
 spherical cap, and Louko and Sorkin \cite{LS} 
have looked at a regularisation based
 on a different coordinate system. Unfortunately things are not quite as
 straightforward as one would hope. It is also possible to choose
 regularisations which do not yield a mass per unit length equal to the
 deficit angle. An important example of this was given by Geroch and
 Traschen \cite{gt}.  They first formed a regularisation sequence ${\tilde
 g}_{(\epsilon)ab}$ which gave the standard answer for the mass per unit
 length. They then introduced a second regularisation sequence ${\hat
 g}_{(\epsilon)ab}$ which was related to the first sequence by a conformal
 factor
 \begin{equation}
 {\hat g}_{(\epsilon)ab}=\Omega^2{\tilde g}_{(\epsilon)ab},
 \end{equation}
 where
 \begin{equation}
 \Omega=\exp(\lambda f(r/\epsilon))
 \end{equation}
 and $f$ is a smooth function whose support is $[1/2,1]$. 
 They then found that the mass per unit length of the limiting 
 spacetime was dependent on both $\lambda$ and $f$:
 \begin{equation}
 \mu=2\pi(1-A)-2\pi\int_{1/2}^1 \lambda^2\sin(\gamma r)f'(r)^2 dr.
 \end{equation} 
 From the point of view of Colombeau theory this ``bad'' behaviour of the
 above regularisation is not unexpected since although $\Omega^2 \approx 1$,
 it is not equal to the unity function in $\G(\R^4)$ and hence ${\hat
 g}_{\epsilon}$ {\em does not} represent the conical spacetime in the full
 Colombeau algebra.  Wilson
 \cite{wilsonDISS} looked at general regularisations of conical spacetimes and
 gave a condition that ensured that a given regularisation sequence gave
 the standard result for the mass per unit length. He showed that provided
 that the difference between the connection one-forms of the regularised
 spacetime and the original spacetime diverged no faster than $1/r$ then
 the answer would be the standard result. Thus although it is possible to
 obtain regularisations which give a different mass per unit length, the
 geometry of these bad regularisations diverges strongly from that of a
 cone as one approaches the axis.
   
 A significant generalisation of the curvature calculations for a 
 conical spacetime was obtained by Wilson \cite{wilson}
 who extended the results to a four dimensional time dependent cosmic
 string. He considered cylindrically symmetric pure radiation solutions
 of Einstein's equations. These metrics may be written in  the form
 \begin{equation}
 ds^2=e^{2\gamma(t-r)}(dt^2-dr^2)+r^2d\phi^2+dz^2
 \end{equation}
 and naively one would expect a delta-function contribution to the
 curvature due to the angular deficit of
 $2\pi(1-e^{-\gamma(t)})$. However, because of the time dependence of the
 angular deficit, the singularity cannot be quasi-regular but must be
 stronger. In fact it is an example of an ``intermediate singularity''
 and the components of the Riemann tensor  have a limit in a special
 choice of frame. Wilson calculated the energy-momentum tensor of this
 spacetime by first writing the metric in null Cartesian coordinates
 and then proceeded to obtain a smooth metric
 by convolution, estimate the curvature of the smooth metric and show
 that it is associated to a distribution. 
 This result shows that even in the radiating case the mass per unit
 length is given by the expected formula
 \begin{equation}
 \mu(u,z) \approx 2\pi(1-e^{-\gamma(u)}).
 \end{equation}
 Furthermore by applying the methods of Ashtekar et al.\ \cite{ash} one may show
 that this result agrees with the asymptotically measured mass per unit
 length.
 However unlike the case of the static string it is unclear whether the
 calculation is coordinate invariant. The $(u,x,y,z)$ coordinates used
 are natural for investigating radiating systems but have the disadvantage 
 that Minkowski space appears to be singular on the axis when written
 in these coordinates. However the correct mass per unit length for a
 static cosmic string is obtained, and this  vanishes for Minkowski
 space. 
 
\section{Nonlinear distributional geometry}

An underlying principle of general relativity is that the measurement
of physical quantities should be independent of the coordinate system
used. Mathematically this is reflected in the fact that the theory
is formulated in terms of tensor fields on manifolds. The minimum
differentiability of the coordinate transformations and the dependence
on the differential structure of the manifold is quite subtle, but at
the very least the theory should be invariant under smooth
diffeomorphisms. 

The definition of the special algebra introduced in \S2
may be generalised in a more or less straight forward manner
to yield a special Colombeau algebra $\G(M)$ on a differentiable manifold $M$ 
\cite{deroever}. This setup was extended by Kunzinger and Steinbauer 
in \cite{KS1} to a theory of generalised sections of vector bundles. 
Furthermore in \cite{KS2} the study of (pseudo-)Riemannian geometry
in the generalised setting was initiated and in \cite{KSV} generalised
connections and curvature in general principal and vector bundles
were studied. However both geodesics and diffeomorphisms involve considering
functions with values in a manifold and if one wishes to consider these
objects in the generalised setting one is forced to consider manifold
valued generalised functions (a concept which it is not possible to deal
with using distributions). The study of such functions was first looked at in
\cite{gfvm}, where the space $\gs[X,Y]$ of generalised functions defined on
the manifold $X$ and taking values in the manifold $Y$ was defined. 
This work was extended to a functorial
theory in \cite{gfvm2} where several global characterisations of the
notions of moderateness and negligibility for generalised functions from $X$
to $Y$ are given. These characterisations provide the key to proving that
composition of generalised functions between manifolds can be carried out 
unrestrictedly. Using these ideas it is possible to consider generalised 
geodesics and flows as well as singular ODEs on manifolds (see
\cite{genflows} for details). An overview over this 
``nonlinear distributional geometry'' can be found in
\cite[Ch. 3.2]{BOOK}.

However in all these constructions the embedding of distributions and
functions of finite differentiability is non-canonical; in addition to the
dependence on the choice of a molifier (see \S2), any embedding into
$\G(M)$ will not be diffeomorphism invariant. So how can this setting be of
use in a diffeomorphism invariant theory like general relativity?

Firstly in some applications there will be a natural physical
parameter, such as a coupling constant, that may be used to directly 
provide a representation of the singular objects involved in $\G$. 
Hence there will be no need for using any embedding at all. 

Another possibility is to model a singular spacetime metric by a whole
class of generalised metrics that is by definition diffeomorphism
invariant.  This has been done in case of the description of impulsive pp
waves in $\G$ (cf \S6).

Finally one may regard the Colombeau algebra as an intermediate
calculational tool for obtaining distributional answers. One picks
some coordinate system, embeds the components of the metric into the
Colombeau algebra (in the given coordinates) and calculates the
components of the curvature and energy-momentum tensor etc. One then
tries to show that these objects are associated to
distributions. Finally one repeats the entire calculation in some
other (smoothly related)  coordinate system and tries to show that the
answer transforms in the expected way for a distribution. 

Although such calculations can be useful in some situations 
it would be preferable to make the embedding into the
algebra coordinate invariant. To explain how this can be done
we first introduce the full Colombeau algebra $\G^e(\R^n)$ 
\cite{colombeau2} which, unlike 
the special version, has a canonical embedding of distributions.

This is achieved by substituting the index set $(0,1]$ by a suitable class of
molifiers. More precisely we introduce the following grading
on the space of all molifiers
\begin{eqnarray*}
{\cal  A}_0  &:=&  \{\Phi  \in  {\cal D}(\R^n) \, : \, \int
\Phi(x)\,dx = 1\}  \\
{\cal  A}_q  &:=&  \{\Phi  \in  {\cal  A}_0 \, : \, \int
\Phi(x)  x^\alpha  \,dx  =  0\,  ,  \,  1\le  |\alpha| \le q\}
\,\,\,(q \in \Nat)
\end{eqnarray*}
and take the basic space to be 
\[\E^e:=\{f:\ {\cal A}_0\times\R^n\to\R^n|\ f\ \mbox{smooth in}\ x\}.\]

Now the respective spaces of moderate and negligible functions 
may be defined as follows (again $\Phi_\eps(x):=(1/\eps^n)\Phi(x/\eps)$).
\bdef
\begin{itemize}
\item[(i)] (Moderate functions)
 \begin{eqnarray*}
{\cal E}^e_M(\R^n)&:=&\{f\in {\cal E}^e:
  \forall K\subset\subset \R^n \ \forall \alpha\in \Nat_0^n \ \exists p\in
  \Nat_0 \ \forall \Phi\in {\cal A}_p:\\  
  &&\qquad \sup\limits_{x\in K} | D^\alpha
  f(\Phi_\eps,x) | =O(\eps^{-p}) \mbox{ as }\eps\to 0\}
  \end{eqnarray*}
 \item[(ii)] (Negligible functions)
 \begin{eqnarray*} {\cal N}^e(\R^n)&:=&\{f\in {\cal E}(\Omega) :
  \forall K\subset\subset \R^n \ \forall \alpha\in \Nat_0^n \ \forall p\in\Nat_0 \
  \exists q \ \forall \Phi\in {\cal A}_q:\\ 
 &&\qquad \sup\limits_{x\in K} | D^\alpha
  R(\Phi_\eps,x) | =O(\eps^p)\mbox{ as }\eps\to 0\}
 \end{eqnarray*}
 \item[(iii)] (Full algebra)
  \[
{\cal G}^e(\R^n):={\cal E}^e_M(\R^n)\,/\,{\cal N}^e(\R^n)\,.
  \]
\end{itemize}
\end{thi}
Distributions are now simply embedded into ${\cal G}^e$ by
convolution with the molifiers i.e.
\begin{equation}
\iota(T)=[T*\Phi]
\end{equation}
and one obtains a differential algebra of generalised functions on $\R^n$
(or open subsets thereof) just as in the special version with the
additional benefit of a canonical embedding of the space of distributions.

Unfortunately this construction cannot be generalised to the manifold
setting in a simple way as the definition of the spaces ${\cal A}_q$ is not
invariant and moreover the embedding is not diffeomorphism invariant since
convolution again depends on the linear structure of $\R^n$.
However an invariant embedding can be achieved by demanding that the 
molifiers $\Phi$ transform in an appropriate way. 
Colombeau and Meril in \cite{CM} made the first decisive steps towards
a diffeomorphism invariant full Colombeau algebra by weakening the moment 
conditions
to only hold asymptotically (which makes them invariant) and enlarging 
$\A_q$ to a space of bounded paths $\epsilon\mapsto\Phi^\epsilon\in\D(\R^n)$. 
A flaw in this
construction was found and removed by Jel\'\i nek \cite{Jel} who 
developed an improved version of the theory which involved some subtle 
changes of definitions and established a number of important technical results.
These ideas were then fully developed and given a firm mathematical basis
in two papers dealing with the foundations of nonlinear
generalised functions  \cite{found} where the first diffeomorphism invariant 
full Colombeau algebra on (open sets of) $\R^n$ was constructed.

For applications in general relativity it is moreover desirable to have
a geometric and global version of the theory rather than simply
giving transformation rules for the local theory. Such a construction was
given in \cite{vim} (for an overview see \cite{annalen}). Here we only
mention that the key idea is to replace the scaled and unbounded paths 
$(1/\epsilon)\Phi^\epsilon$ which are employed in the definition of 
moderateness and negligibility in the local theory by {\it smoothing kernels}
$\Phi$ which are $C^\infty$ maps from $(0,1]\times M$ to compactly supported
$n$-forms on $M$. In this way one obtains a geometrically constructed full
Colombeau algebra on a differential manifold $M$ where the canonical embedding
of distributions commutes with Lie derivatives.

However this theory still lacks a canonical embedding of distributional 
{\it tensors}.  Although the work of \cite{vim} described above provides 
one with an
invariant embedding of scalars, one cannot simply apply this to the 
components of a tensor and obtain an invariant embedding of the tensor. 
Embedding the components of a tensor and then transforming
would in general give a different answer from transforming and then
embedding since multiplication by a smooth function does not in general
commute with the embedding. The problem really stems from the way the
convolution integrates the components of the tensor at {\it different}
points of the manifold. The solution to this is to introduce some
additional structure which enables one to first transport the tensor
fields to the same point $p$ in $M$ so that one can then do the
integration in a meaningful way. Such a transport is
naturally provided by specifying a background connection, 
and in keeping with the
spirit of the full algebra this is made an argument of the generalised
tensor field. Thus a generalised tensor field depends upon a
background connection $\gamma$, a smoothing kernel $\Phi$ and the
point $p$. This is currently work in progress but see \cite{sotonTF} and \cite{new} for a more detailed description of this approach.

This approach now provides a "nonlinear distributional geometry" plus
a canonical and invariant embedding of distributional objects.
For example one can canonically embed the conical metric of the
previous section to obtain a generalised metric. One then finds that
this metric has a curvature tensor which is associated to the Dirac
distribution and this shows that in any
coordinate system a conical spacetime has a generalised Einstein
tensor which satisfies
\begin{equation}
\tilde G_{ab} \approx \tilde T_{ab}
\end{equation}
where $\tilde T_{ab}$ is the embedding into the Colombeau algebra of
the energy-momentum tensor of a (thin) cosmic string.

\section{Weak singularities and Generalised hyperbolicity}          

As discussed earlier the definition of a singularity in general relativity
is different from other field theories where one has a background
metric. In general relativity one
detects the presence of
singularities by showing that the spacetime is incomplete in some
sense. In the standard approach to singularities (see e.g.\ Hawking and
Ellis, \cite{hawking+ellis}), a singularity is regarded as an 
obstruction to extending a
geodesic. However this definition does not correspond very closely to ones
physical intuition of a singularity. This led to a consideration of whether
physical objects would be subjected to unbounded deformations as one
approached the singularity and was formulated mathematically in terms of
strong curvature conditions. Unfortunately it is hard to model the
behaviour of real physical objects in a strong gravitational field and
because of this Clarke \cite{clarke} suggested that one should consider instead
the behaviour of physical fields (for which one has a precise mathematical
description) near the singularity. According to the philosophy of
``generalised hyperbolicity'' one should regard singularities as obstructions
to the Cauchy development of these fields rather than an obstruction to the
extension of geodesics.  However even a mild singularity is an
obstruction if one uses the standard theory of distributions. By
considering solutions to the wave equation in the Colombeau algebra
one finds a certain class of weak singularities which 
do not prevent the evolution of test fields (Vickers and Wilson, \cite{VW1}).

In this section we will look at solving the wave equation on a spacetime
with a locally bounded singular metric 
(such as the conical spacetime considered in  \S7). 
The standard Cauchy problem takes the form 
\begin{eqnarray}
  \square u(t,x^\alpha) &=& 0  \\
  u(0,x^\alpha) &=& v(x^\alpha) \nonumber \\
  {\partial_t u}(0,x^\alpha) &=& w(x^\alpha) \nonumber 
\end{eqnarray}
with initial data $(v,w)$ lying in the Sobolev spaces  
$H^1(S)\times H^0(S)$ prescribed on the
initial surface $S$, given by $t=0$.  
Because of the form of the metric one would expect the solution (if it
exists) to be defined as a distribution. 
This however will cause difficulty in interpreting
\begin{equation} 
\square u = 
(-g)^{-1/2} \partial_a \bigl( (-g)^{1/2} g^{ab} \partial_b u \bigr) 
\end{equation}   
as a distribution in the framework of classical distribution theory
because the above equation has (non-constant) singular coefficients and 
involves ill-defined products. 
Again we overcome these difficulties by using
the nonlinear generalised function theory of Colombeau.  
For an overview of the treatment of PDEs with singular coefficients, data
and solutions in this setting see \cite{MO}.

We first canonically embed the metric $g_{ab}$ into the
full Colombeau algebra $\G^e(M)$ by using a convolution integral
(\ref{smoothmetric}) as in \S7 to obtain a representative 
$(\tilde{g}_{(\epsilon)ab})\in\E_M^e(M)$. 
Since the initial data $(v,w)$ does not have to be smooth, we must also
embed it into the algebra as $(V,W)$ represented again by convolution 
integrals denoted $\veps$ and $\weps$ respectively.

The generalised function wave operator acting on a generalised function 
$U$ represented by $u_\epsilon$ may then be written as
\begin{equation} \boxeps \ueps = (-\geps)^{-1/2}
    \partial_a\bigl( (-\geps)^{1/2} \gueps{ab} \partial_b \ueps \bigr).
\end{equation}
We would like to then be able to solve the  Cauchy problem in the space
$\G^e(M)$
\begin{eqnarray}
   \square U(t,x^\alpha) &=& 0  \label{cauchy1}\\
   U(0,x^\alpha) &=& V(x^\alpha) \nonumber \\
   {\partial_t U}(0,x^\alpha) &=& W(x^\alpha) \nonumber
\end{eqnarray}
and obtain a solution $U\in\G^e(M)$ which is associated to a distribution.
In practice one works with the equivalent problem in $\E_M^e(M)$
\begin{eqnarray}
   \boxeps \ueps(t,x^\alpha)  &=& \feps(t,x^\alpha)  \label{cauchy2} \\
   \ueps(0,x^\alpha) &=& \veps(x^\alpha)  \nonumber \\
   {\partial_t\ueps}(0,x^\alpha) &=& \weps(x^\alpha) \nonumber
\end{eqnarray}
where $(\feps)$ is negligible.

%
%

In \cite{VW1} it is shown how to estimate solutions $\ueps$ of 
(\ref{cauchy2}) and its derivatives in terms of powers of $\eps$, 
given the moderate and null bounds of $\feps$, $\veps$ and $\weps$, 
using a method of energy estimates following Hawking and Ellis 
\cite{hawking+ellis} 
and Clarke \cite{clarke}. The additional complication in this situation 
is that one needs more explicit bounds because one needs to know the
precise way in which the constants depend upon $\epsilon$. This is
accomplished by working with function spaces defined using higher order
energy estimates (related to the super-energy tensors of Senovilla
\cite{seno}) for which one has bounds in terms of the covariant 
derivatives of the curvature. Using this method one can show that the
Cauchy problem (\ref{cauchy1}) has a unique solution $U \in
\G^e(M)$ which is independent of the representation chosen for
$g_{ab}$, $V$ or $W$. In the case of a conical spacetime one can go further
and show that this solution is actually associated to a 
distributional solution.
Thus a conical spacetime satisfies the condition of ``generalised
hyperbolicity'' as claimed. It seems likely that these techniques may then 
be used to show that a much wider class of spacetimes with weak singularities
satisfy the generalised hyperbolicity condition. See (\cite{mayerhofer}) 
for some recent results in this direction.

\section{Conclusion}

In this review we have looked at the extent to which it is possible to use
conventional distribution theory to look at solutions of Einstein's
equations. Although there is an important class of new solutions that can
be obtained by going beyond the confines of $C^{2-}$ metrics the largest
class that one can work with that is ``stable'' is given by the gt-regular
metrics. Such metrics can be used to describe solutions with singular
support of the curvature on a hypersurface but are
unable to deal with singularities of higher codimension. To deal with these
it is necessary to go beyond distributions and work with a theory of
nonlinear generalised functions. We have shown that such an appropriate
description of nonlinear generalised functions is given by the theory of
Colombeau algebras. 

The special theory provides a straightforward
computational tool for calculating the distributional curvature of a number
of singular metrics and throwing some light on the physical nature of the
singularity. Furthermore it has also been possible to define generalised
functions taking values in a manifold and this allows one to talk about
generalised geodesics and generalised symmetries (see e.g.\ \cite{ab2}) 
of a spacetime. Unfortunately due to lack of space we have had to omit from this review
both this latter topic and the topic of impulsive pp-waves
and ultrarelativistic black holes with non-vanishing cosmological
constant (see e.g.\ \cite{jiri-iwid}). 

However the special algebra does not provide one with a
canonical embedding of distributions so there is always a question about
the extent to which the answer depends upon the particular embedding that
is used. The full Colombeau algebra rectifies this problem and a global
formulation that is independent of the coordinate system has been
given. Although
the details of the tensorial  theory remain to be fully worked out this 
work provides the basis for a coordinate and embedding independent
theory of generalised (pseudo-)Riemannian geometry which can be used to
analyse a wide class of singular spacetimes. In particular it is possible
to give a distributional interpretation to a many physically
reasonable singularities. The remaining singularities can therefore be
regarded as true gravitational singularities. An outstanding
project is to consider the singularity theorems in this generalised 
setting and show that they predict true gravitational singularities rather
than simply distributional singularities.

\ack
We acknowledge support by Austrian Science Fund (FWF) grants P16742 and Y237.

\end{document}